\documentclass[twocolumn]{aastex62}

\graphicspath{{./}{figures/}}

\received{XXX}
\revised{YYY}
\accepted{ZZZ}
\submitjournal{ApJ}

\usepackage{graphicx}	
\usepackage{amsmath}	
\usepackage{amssymb}	
\usepackage[inline]{enumitem}
\usepackage{gensymb}

\usepackage{color}

\usepackage{xcolor}

\shorttitle{Giant Planets, Tiny Stars}
\shortauthors{Stephan et al. 2021}

\begin{document}

\title{Giant Planets, Tiny Stars:\\Producing Short-Period Planets around White Dwarfs with the Eccentric Kozai-Lidov Mechanism}

\correspondingauthor{Alexander P. Stephan}
\email{stephan.98@osu.edu}

\author[0000-0001-8220-0548]{Alexander P. Stephan}
\affiliation{Department of Astronomy, The Ohio State University, Columbus, OH 43210, USA}
\affil{Department of Physics and Astronomy, University of California, Los Angeles, Los Angeles, CA 90095, USA}
\affiliation{Mani L. Bhaumik Institute for Theoretical Physics, University of California, Los Angeles, Los Angeles, CA 90095, USA}

\author[0000-0002-9802-9279]{Smadar Naoz}
\affiliation{Department of Physics and Astronomy, University of California, Los Angeles, Los Angeles, CA 90095, USA}
\affiliation{Mani L. Bhaumik Institute for Theoretical Physics, University of California, Los Angeles, Los Angeles, CA 90095, USA}

\author[0000-0003-0395-9869]{B. Scott Gaudi}
\affiliation{Department of Astronomy, The Ohio State University, Columbus, OH 43210, USA}



\begin{abstract}
The recent discoveries of WD J091405.30+191412.25 (WD J0914 hereafter), a white dwarf likely accreting material from an ice giant planet, and WD 1856+534 b (WD 1856 b hereafter), a Jupiter-sized planet transiting a white dwarf, are the first direct evidence of giant planets orbiting white dwarfs. However, for both systems the observations indicate that the planets' current orbital distances would have put them inside the stellar envelope during the red giant phase, implying that the planets must have migrated to their current orbits after their host stars became white dwarfs. Furthermore, WD J0914 is a very hot white dwarf with a short cooling time that indicates a fast migration mechanism. Here, we demonstrate that the Eccentric Kozai-Lidov (EKL) Mechanism, combined with stellar evolution and tidal effects, can naturally produce the observed orbital configurations, assuming that the white dwarfs have distant stellar companions. Indeed, WD 1856 is part of a stellar triple system, being a distant companion to a stellar binary. We provide constraints for the orbital and physical characteristics for the potential stellar companion of WD J0914 and determine the initial orbital parameters of the WD 1856 system.

    
\end{abstract}


\section{Introduction}\label{sec:intro}
 
White dwarfs (WDs), the last life stage and remnants of most stars ($\leq8$~M{$_\odot$}), have ended all nuclear fusion processes and should have atmospheres made purely out of hydrogen and helium, as heavier elements settle into the WD's core \citep{Paquette+1986_A}. However, observations over the last few decades have revealed that a large fraction of WDs show significant amounts of heavy elements in their atmospheres, often consistent with the bulk composition and amounts of rocky asteroids or terrestrial planets \citep[e.g.,][]{DebesSigurdsson2002,Jura2003, Jura+2009,Zuckerman+2003,Zuckerman+2007,Zuckerman+2010,Zuckerman+2011,Veras+2017_B}, which has lead to the hypothesis that planetary material pollutes WDs. Indeed, further discoveries over recent years have given support to this idea, as disintegrating planetary bodies \citep[e.g.,][]{Vanderburg+2015,Xu+2016} and debris disks \citep[e.g.,][]{Koester+2014,Farihi2016} have been observed around WDs. Many different dynamical processes have been suggested to explain how planetary material can be brought onto a WD, such as planet-planet scattering \citep[e.g.,][]{Bonsor+2011, Frewen+2014, Mustill+2014, Mustill+2018}, orbital instabilities \citep[e.g.,][]{Veras+2013, BonsorVeras2015}, and secular effects such as the Eccentric Kozai-Lidov \citep[EKL, e.g.,][]{Naoz2016} mechanism in hierarchical three-body systems \citep[e.g.,][]{HPZ2016,PetrovichMunoz2017,Stephan+2017}. The EKL mechanism is particularly well-suited to explain the observed nitrogen abundance in WD 1425+540, a WD with a distant stellar companion that was most likely polluted by long-period comet-like bodies from a Kuiper-belt analog, delivering volatile-rich icy material to the WD \citep{Xu+2017, Stephan+2017}. Indeed, as was shown in \citet{Stephan+2017, Stephan+2018}, the EKL mechanism can not only cause the migration of such comets, but can also cause giant planets to pollute WDs by increasing their orbital eccentricities to values of $e\sim1$, feeding material to the WDs.

The recent discovery of the polluted white dwarf WD J091405.30+191412.25 (WD J0914, for simplicity) by \citet{Gaensike+2019} appears to confirm the prediction of \citet{Stephan+2017, Stephan+2018}, as WD J0914 has been found to accrete material from an ice giant planet, estimated to be on a close (about $0.07$~AU) orbit. The planet must have migrated to its current orbit after the WD formed, in order to avoid engulfment and destruction during the star's red giant phase. Interestingly, the cooling age of the WD is extremely short, estimated at only about $13$~Myrs, requiring a very rapid migration mechanism. That mechanism has been suggested to be high eccentricity migration involving chaotic tides \citep[e.g.,][]{VerasFuller2019a}. 

Another recent discovery, WD 1856+534 b (WD 1856 b, for simplicity), is an even stronger candidate for a giant planet orbiting a WD, as it was discovered through the transit method \citep{Vanderburg+2020}. This planet orbits its WD host on an even shorter orbit than that estimated for WD J0914, at only about $0.02$~AU.

In this paper we show how the EKL mechanism can naturally explain both of these discoveries, as was predicted by the model of \citet{Stephan+2017}, by inducing the large eccentricities required for tidal migration to occur. A crucial element for this mechanism is the enhancement of the EKL mechanism through the mass loss occurring when the primary star becomes a WD, which was first discussed for triple star systems as the ``mass-loss-induced eccentric Kozai'' in \citet{Shappee+13} \citep[see also][]{Michaely+14}. The mass loss changes the orbital configuration of a given system, such that orbital architectures that were not undergoing strong EKL oscillations during the main sequence lifetime of the primary star might begin to do so once the WD has formed. While the Kozai-Lidov (KL) effect \citep{Kozai,Lidov,Naoz2016} has been discussed in recent papers as a possible explanation for WD 1856 b, the qualitative change in the dynamical evolution due to the ``mass-loss-induced eccentric Kozai'', in particular the ``triggering'' of inclination flips and associated high-eccentricity spikes that significantly change the dynamical behavior of a system as its host star is evolving past the main sequence, was generally not the main focus of those works \citep{OConnor+2020,Munoz+2020}.

We perform Monte-Carlo simulations to determine the original system parameters best suited to reproduce WD J0914 and WD 1856. The results provide constraints for the initial orbital parameters of the planets and for the hypothetical stellar binary companion to WD J0914, which will be useful for further observations and studies of the system.

\section{Observational Constraints}\label{sec:ObsCon}

\subsection{WD J0914 Observations}\label{sec:constraints}

While the original orbital parameters of the ice giant planet orbiting WD J0914 are not known, the observations by \citet{Gaensike+2019} provide some important constraints. The current semi-major axis (SMA) is estimated to be on the order of $15$~R$_\odot$ ($0.07$~AU), however, the orbital eccentricity is unknown. The gaseous circumstellar disk extends to about $10$~R$_\odot$ ($0.045$~AU), indicating that the planet's pericenter is further out than that distance. The original mass of the star during its main sequence lifetime is inferred to have been between $1$ and $1.6$~M$_\odot$, with a current mass of $0.56$~M$_\odot$. As has been pointed out by \citet{VerasFuller2019b}, in order to survive the red giant phase the planet must have been at least $2$ to $3$~AU away from the star originally. However, ice giant planets, or giant planets in general, are known to exist with orbits as large as several tens of AU around many stars, such as in our own solar system or in HR 8799 \citep{Marois+2008,Marois+10}. Thus, the original orbital SMA of the planet is poorly constrained, as is its eccentricity, apart from the estimated current distance. 

\begin{figure*}
\hspace{0.0\linewidth}
\includegraphics[width=\linewidth]{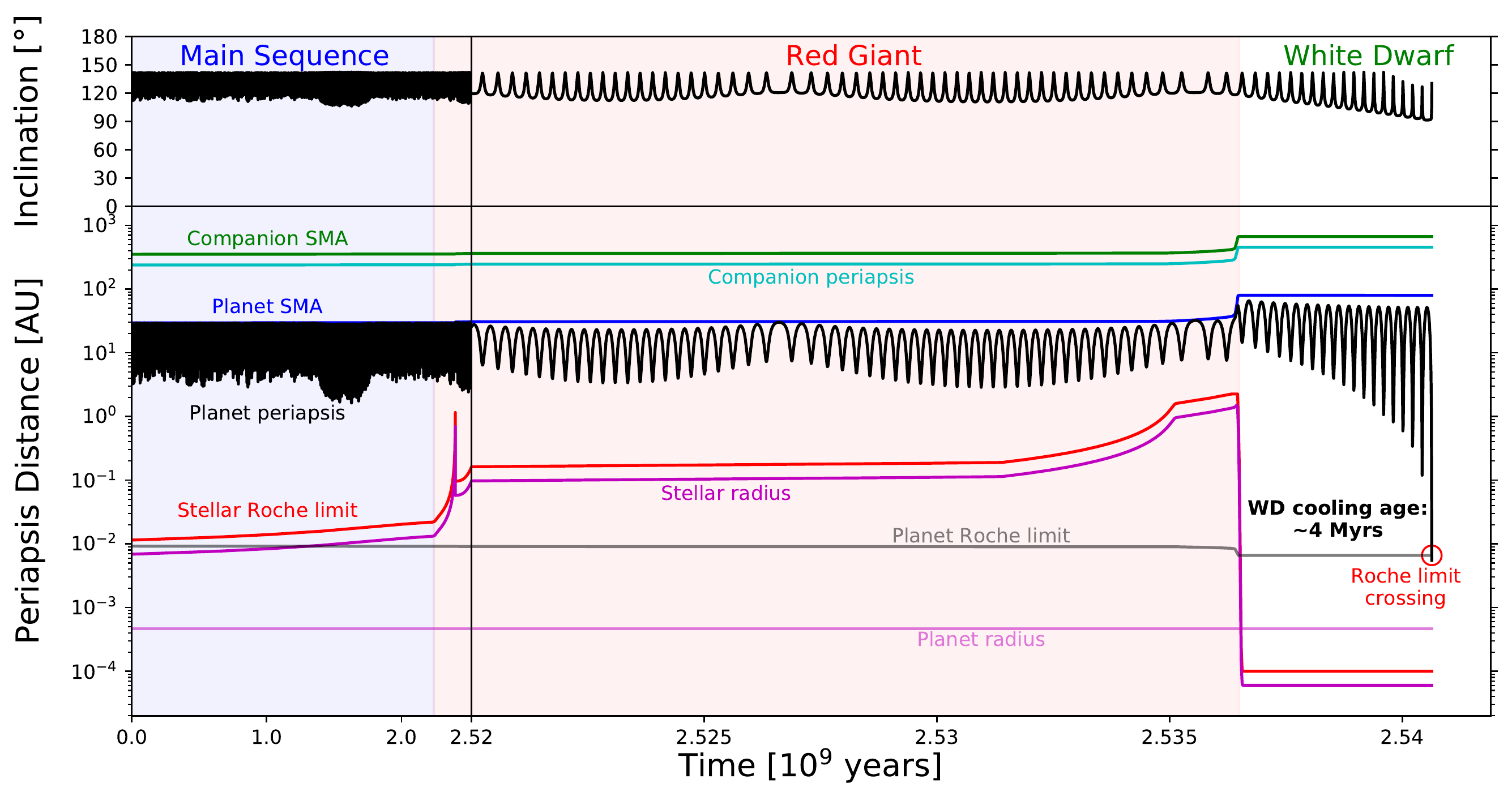}
\caption{{\bf Example orbital evolution of a WD J0914-like progenitor system.} This figure shows the orbital evolution of a Jupiter-like planet orbiting the larger star in a stellar binary system. The planet is eventually tidally disrupted $4$ Myrs after its host star has become a WD, potentially leaving behind a remnant planet similar to the one observed in the WD J0914 system. {We note here that we consider a tidal disruption event sufficient to explain the observations of the WD J0914 system, as no planet is directly observed there, only gas that stems from a potential planet (see also our list of simulation outcomes in Section \ref{sec:setup}).} The upper panels show the system's inclination evolution, the lower panels show the evolution of the planet's periapsis (black) and semi-major axis (SMA) (blue), the stellar companion's periapsis (cyan) and SMA (green), the host star's radius (magenta) and Roche limit (red), and the planet's radius (light magenta) and Roche limit (grey). The red circle marks the time when the planet's periapsis brings it so close to the star that the planet is disrupted. The host star's main sequence, red giant, and WD phases are marked by the blue, red, and white backgrounds, respectively. The left panels show the evolution for the first $2.52$ Gyrs, while the right panels zoom in onto the last $\sim20$ Myrs before disruption. Note how the moderate KL oscillations during the main sequence and red giant phases increase in strength after the main star has lost mass and has become a WD after about $2.5365$ Gyrs of evolution. {\bf Initial conditions:} $m_1=1.6$~M$_\odot$, $m_p=M_{Jup}$ $m_2=0.534$~M$_\odot$, $R_p=R_{Jup}$, $a_{in}=29.5$~AU, $a_{out}=352$~AU, $e_{in}=0.01$, $e_{out}=0.322$, $inc=120.3\degree$, $g_{in}=13.65\degree$, and $g_{out}=57.41\degree$.}
\label{fig:Example2020}
\end{figure*}

Another, much stronger, constraint is given by the cooling age of the WD, which has been determined to be about $13$~Myrs \citep{Gaensike+2019}. While the total age of the star is relatively uncertain, on the order of hundreds of Myrs to several Gyrs, due to the range of possible progenitor masses, the time since the star has become a WD can be determined rather precisely. Therefore, any proposed dynamical mechanism that aims to explain the change in the planet's orbit must be able to do so within a time span of a few Myrs after the formation of the WD. Furthermore, the planet itself should be of rather low density, in order for chaotic tides to be able to shrink the orbit sufficiently over such a short time span and at the estimated current SMA distance \citep{VerasFuller2019b}. 

These observations and previous results imply that the planet currently polluting WD J0914 must have remained on an orbit beyond $2$ or $3$~AU during the main sequence and red giant phases of the host star, but must have migrated to its current orbit on a timescale on the order of $10$~Myrs once the star became a WD. The evolutionary model we explore here must follow these constraints.

\subsection{WD 1856 Observations}\label{sec:constraints2}

The physical characteristics of WD 1856 and its planet candidate are much better constrained than for the WD J0914 system, as WD 1856 b was discovered from its transit of the host WD using NASA's Transiting Exoplanet Survey Satellite (TESS) \citep{Vanderburg+2020}. The orbital period of the transiting planet is $\sim1.4$~days. The current mass of the WD is about $0.52$~M$_\odot$, indicating a SMA distance of $0.02$~AU. Its orbital eccentricity is not well constrained, and may still be substantial (potentially up to $0.68$). The estimated cooling age of the WD is quite old, at about $6$~Gyrs, with a total system age of about $10$~Gyrs, which provides limits for the maximum possible mass of the planetary candidate. As such, the planet is estimated to not be more massive than $11.7$~M$_{Jup}$, {as it would otherwise produce a strong enough thermal radiation signal to have been detected by \citet{Vanderburg+2020}}. Like for the WD J0914 system, the planet could not have survived at its current orbital distance during the main sequence or red giant phases of the host star, but must have originated on a much wider orbit. The original mass and radius of the host star, however, is not well constrained, as is pointed out by \citet{Vanderburg+2020}. Nevertheless, given the limits set by the age of the universe and the cooling age of the WD, it is reasonable to assume that the WD progenitor mass was somewhere in the range of $1.3$ to $1.6$~M$_\odot$. As such, the planet must have originated on an orbit beyond about $2$~AU to safely survive the red giant phase of the primary.

Beyond the detailed orbital information obtained by the transit data, we also know that WD 1856 is part of a visual triple star system. The stellar binary pair G 229-20 A/B has a visual separation from WD 1856 of about $1000$~AU, with a projected binary separation of about $56$~AU. G 229-20 A/B are two M-dwarfs with nearly equal masses, with a combined mass of roughly $0.65$~M$_\odot$. Based on analysis by \citet{Vanderburg+2020}, the binary pair and the WD orbit each other with a SMA of about $1500^{+700}_{-200}$~AU and with some modest eccentricity ($0.30^{+0.19}_{-0.10}$). The existence of these companions provides limits for the original orbital architecture of the system. {Given the preliminary nature of the orbital analysis for WD 1856, we take the visual separation of $1000$~AU as a lower limit for the current SMA, while we set the largest estimate of $2200$~AU as the upper limit. The estimated original combined mass of the WD progenitor and companion binary is in the range of $1.95$ to $2.25$~M$_\odot$. Given the current combined mass of the WD and companion binary of about $1.17$~M$_\odot$, the mass of the system decreased by a factor of about $1.66$ to $2$. Assuming adiabatic mass loss, this decrease in mass should be directly proportional to the expansion of the SMA, and the original SMA between the WD progenitor and the M-dwarf binary should therefore have been in the range of $500$ to $1100$~AU}. From the perspective of the planet, G 229-20 A/B mostly act as a third perturber \citep[ignoring additional dynamical effects as discussed by][]{OConnor+2020}, thus making this system a hierarchical three-body system that can undergo EKL oscillations. As we discuss in Section \ref{sec:EKL}, the EKL mechanism can enable distant planets to undergo high-eccentricity tidal migration, possibly producing the currently observed system configuration. We note here that the binarity of G 229-20 A/B may have an influence on some of the orbital configurations investigated here; however, as previous works \citep[e.g.,][]{Pejcha+13, HamersZwart2016, HamersLai2017} have shown, the result is generally an enhancement of EKL effects, causing more frequent and intense eccentricity spikes, depending on the SMA ratios within the system. We have thus chosen to ignore these effects to reach more conservative conclusions.

\section{The EKL Mechanism and Numerical Methods}\label{sec:EKL}

The EKL mechanism can induce extreme values for the orbital eccentricity $e$ in hierarchical three-body systems, often such that $1-e<10^{-4}$, bringing the orbiting bodies into close proximity or even contact during pericenter passage \citep[e.g.,][]{Naoz11,Naoz+11sec,Naoz+12bin,Li+13,Li+14}. However, if such eccentricities are reached during the star's main sequence or red giant phase, an orbiting body such as a planet might be engulfed and consumed by its host star, preventing it from polluting the resulting WD \citep[e.g.,][]{Stephan+2018,Stephan+2020}. However, as was shown in \citet{Stephan+2017}, the mass loss a star undergoes as it becomes a WD can fundamentally change the dynamical behavior of a hierarchical star-planet-star system, often increasing the strength of EKL oscillations, such that planets that would not come close to their host stars during earlier life stages will actually do so once the star has become a WD, causing pollution or enabling high-eccentricity tidal migration. In this work we follow the same basic model as used in \citet{Stephan+2017}, which we briefly explain here \citep[see also][for the mechanism in the context of triple stars]{Shappee+13,Michaely+14}.

\begin{figure*}
\hspace{0.0\linewidth}
\includegraphics[width=\linewidth]{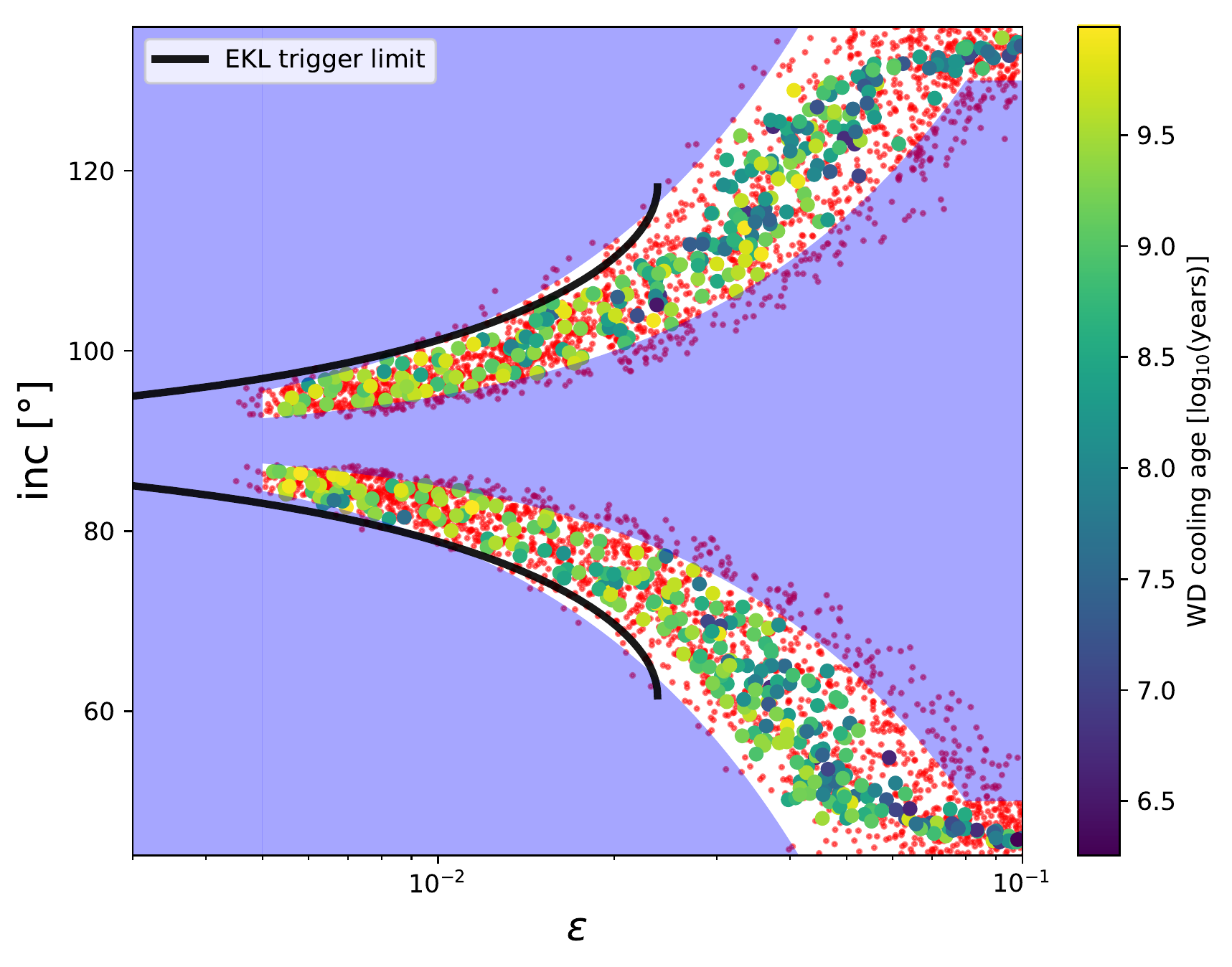}
\caption{{\bf Parameter space relevant for mass-loss triggered EKL.} {The figure shows the $\epsilon$-inclination parameter space for our simulations. For a system to undergo mass-loss triggered EKL evolution, it initially has to be positioned to the left of the theoretical EKL-flip limit \citep[black line, based on, for example,][]{Boaz2} and be moved to the right of it as the value of $\epsilon$ increases as the main star loses mass. We chose systems with initial conditions that are close to this transition zone, with the red dots showing their predicted final position in the parameter space at the end of stellar mass loss. The yellow, green, and blue dots highlight systems that actually underwent WD pollution or planet migration in our simulations, with the colors representing the WD cooling age at the time of pollution or migration (yellow dots representing older, blue dots younger WDs, as shown by the color bar to the right). The blue shaded regions mark parts of the parameter space where the mass-loss triggered EKL mechanism is not relevant, based on our results and previous work \citep{Stephan+2017}, as systems to the right of the EKL-flip limit and close to $90\degree$ inclination can undergo EKL already during the host star's main sequence, while systems to the left of the limit will never undergo EKL.}}
\label{fig:Paramspace}
\end{figure*}

We use the hierarchical secular dynamical evolution equations (which describe the EKL mechanism) derived from solving the hierarchical three-body Hamiltonian up to the octupole order of approximation and averaging over the orbits \citep[e.g.,][]{Naoz+11sec,Naoz2016}. In hierarchical three-body systems, two bodies orbit each other relatively closely as an {\it inner binary}, in this case the WD progenitor and the planet, and are in turn orbited by a distant body, for example the hypothesized stellar companion for WD J0914 or the known M-dwarf binary companions for WD 1856, forming an {\it outer binary}. We assume that the M-dwarf binary companion to WD 1856 mostly acts as a single object from the perspective of the inner binary, however, under certain conditions, the companion binary may enhance the standard EKL mechanism through secular four-body interactions \citep[e.g.,][]{Pejcha+13,Hamers+2015,HamersPZ2016,HamersLai2017}. The enhancements may have contributed to the formation of WD 1856, if the planet's initial orbit was wide enough \citep{OConnor+2020}. Additionally, we include tidal effects for the inner binary based on the equilibrium tide model of \citet{Hut} and \citet{1998KEM}, general relativistic precession based on \citet{Naoz+12GR}, as well as stellar evolution effects for stellar radii, masses, and structure changes based on the stellar evolution code {\tt SSE} by \citet{Hurley+00}, with adjustments to the magnetic braking coefficients based on \citet{DD+2004} and \citet{BarkerOgilvie2009}. Also see \citet{Naoz2016} for the complete set of dynamical equations. 

Using these methods we construct a set of simulations to explore the WD J0914 and WD 1856 system evolution over time and determine their likely initial system parameters, as well as the likelihood that the observed outcomes are the result of EKL evolution.

\subsection{Initial Conditions}\label{sec:setup}

To explore the orbital configurations that can produce systems like WD J0914 or WD 1856, we perform a Monte-Carlo simulation of $4000$ systems with the following initial conditions: The initial mass of the primary star, $m_1$, is chosen uniformly between $1$ and {$1.6$~M$_\odot$}, based on the estimated initial mass by \citet{Gaensike+2019} and reasonable assumptions for the mass based on \citet{Vanderburg+2020}, with its radius, $R_1$, and initial spin period, $P_{spin,1}$ determined by {\tt SSE}. The mass, radius, and spin periods of the planet, $m_p$, $R_p$, and $P_{spin,p}$, respectively, are set arbitrarily to the mass, radius, and spin period of either Neptune or Jupiter. The distant stellar companion's mass, $m_2$, is based on observations of field binary mass ratios by \citet{Duquennoy+91}. The planet's SMA, $a_{in}$, is chosen uniformly between $3$ and $100$~AU, and its initial eccentricity, $e_{in}$, is set close to zero ($e_{in}=0.01$). The outer companion star's SMA, $a_{out}$, is chosen from a log-normal distribution with a peak at a few dozen AU, as observed for field binaries by \citet{Duquennoy+91} \citep[see also][for more recent studies with broadly consistent results]{Raghavan+10, Moe+2017}, and its eccentricity, $e_{out}$, is chosen uniformly between $0$ and $1$. However, orbital stability and hierarchical system configuration require that the factor $\epsilon<0.1$ and that $a_{in}/a_{out}<0.1$ \citep[see][]{Naoz2016}. Here $\epsilon$ is defined as \begin{equation}\label{eq:eps}
    \epsilon = \frac{a_{in}}{a_{out}}\frac{e_{out}}{1-e_{out}^2} \ ,
\end{equation} 
\citep[e.g.,][]{LN}\footnote{{We use the hierarchy requirements as our stability criteria as they are numerically consistent with stability, especially for hierarchical systems \citep[see also][]{Bhaskar+2021}.}}. 
These restrictions limit the possible stellar companion SMA values to be larger than about $30$~AU to $1000$~AU, depending on the initial planetary orbit, and disfavor very large eccentricities for the stellar companion's orbit. Furthermore, stellar binary orbits larger than $10,000$~AU are also excluded, as Galactic tides can unbind those systems over relatively short timescales. The initial inner and outer arguments of periapsis ($g_{in}$ and $g_{out}$) are chosen uniformly between $0\degree$ and $360\degree$. The initial inclination, $inc$, between the inner and outer orbits is chosen uniformly in cosine, however, it is limited between $40\degree$ and $140\degree$, as smaller or larger inclination values will generally not lead to EKL oscillations.

Additionally, we consider another constraint for the applicable parameter space for $\epsilon$ (see Eq. \ref{eq:eps}). As mentioned above, \citet{Stephan+2017} showed that stellar mass loss can trigger stronger EKL oscillations, in particular by enabling octupole-level effects of the EKL mechanism to drive extreme eccentricity spikes and orbital inclination flips (see Fig. \ref{fig:Example2020} for an example). The shift in dynamical behavior is caused by changes to the value of $\epsilon$ due to the mass loss of the primary star, which expands the orbital SMAs within the system. Specifically, in a system where the primary object and the outer companion are much more massive than the inner companion, the SMA ratio between the inner and outer orbits will change, leading to a growth in $\epsilon$. The value of $\epsilon$ after mass loss, $\epsilon_f$, can be described by \begin{equation}\label{eq:epsf}
    \epsilon_f = \frac{m_{1,i}}{m_{1,f}}\frac{m_{1,f}+m_2}{m_{1,i}+m_2} \epsilon_i,
\end{equation} where the subscripts $i$ and $f$ denote the values of the primary star's mass, $m_1$, and $\epsilon$ before and after mass loss, respectively, the outer companion mass, $m_2$, is assumed to remain constant, and the planet is treated as a test particle. {We note here that due to this mechanism even a small increase in the value of $\epsilon$, even by less than a factor of two, can completely alter the dynamical behavior of the planet.} However, the value of $\epsilon$ does not determine if octupole-level effects are active by itself, but does so in conjunction with the value of the inclination. We thus restricted our simulations to systems that are close to the theoretically and experimentally determined threshold in the $\epsilon$-inclination parameter space, as we show in Fig.~\ref{fig:Paramspace} \citep[see also Fig.~2 in][]{Stephan+2017}. {In particular, we only considered systems that are initially in the non-EKL-active part of the $\epsilon$-inclination parameter space and are predicted to cross into the EKL-active part of that space due to mass loss induced increases in the value of $\epsilon$. The rest of the parameter space, highlighted by the blue background in Fig.~\ref{fig:Paramspace}, is excluded as those systems would either already undergo EKL during the host star's main sequence, or never undergo EKl even after mass loss has occurred.}

Using these initial conditions, the systems are evolved through all stellar evolution phases for up to $13$~Gyrs, or until the planet comes in contact with the stellar surface and is destroyed. If a planet survives to the host star's WD phase, it can undergo three different fates:\begin{enumerate}
    \item The strength of the EKL oscillations do not increase and the planet remains on a wide orbit.
    \item The strength of EKL increases enough to bring the periastron distance close to the WD, enabling tidal interactions to reduce the SMA over relatively short timescales \citep[estimated to be closer than about $0.02$~AU following][unless the planet is of very low density]{VerasFuller2019a}.
    \item The strength of EKL increases so much that the planet's periastron is driven inside the Roche limit, implying tidal stripping of the planet by the WD.
\end{enumerate} Case (2) is the most direct example for forming a WD J0914 or WD 1856-like system, however, case (3) could also result in such a system, with the currently observed planets being a remnant of a larger progenitor. {This could especially be the case for WD J0914, as the planet is not directly observed, only gas that may have been deposited through tidal disruption. Additionally, the planets in neither WD J0914 nor WD 1856 need to have finished circularizing based on current observations. We note that, going beyond the equilibrium tides model used here, to include tidal models that involve the structure of the planet, such as chaotic tides, would probably increase the efficiency of inward migration \citep[e.g.,][]{Vick+2019}. This increased efficiency may also increase the risk of planet disruption \citep[e.g.,][]{Naoz+12bin,Petrovich2015,Anderson+2016}.} In section \ref{sec:results} we show the initial system configurations most likely to produce outcomes (2) and (3). As WD J0914 has a cooling age of about $13$~Myrs, only systems that reach outcomes (2) or (3) on similar timescales are counted as reproducing the WD J0914 system. For WD 1856, the long cooling time does not provide a similar restriction, however, as it has known companions, only systems with outer eccentricities, post-mass loss SMAs, and stellar masses close to the likely estimated ranges are considered in our analysis of the results (here we assume $0.2<e_{out}<0.5$, $1000$~AU $<a_{out, final}<2500$~AU, and $0.5$~M$_\odot < m_2 < 0.8$~M$_\odot$, to take various measurement uncertainties for WD 1856 into account).

\section{Results}\label{sec:results}

From the results of our $4000$ simulated systems we are able to deduce the parameters and approximate occurrence rates for producing WD J0914 and WD 1856-like systems. In about $14~\%$ of simulated systems the planet either came close enough to the star to undergo tidal stripping or rapid tidally induced orbital shrinking such that it reached SMA values closer than about $0.02$~AU. Given our restricted parameter space, these $14~\%$ translate to an overall rate on the order of one of a percent or less, which is broadly consistent with the observations given that only two systems have been discovered so far among hundreds of known polluted WDs. As can be seen in Fig. \ref{fig:Paramspace}, the WD cooling age at which a particular planet can reach such a short-period configuration is broadly a function of the initial inclination and the final $\epsilon$ value. As such, systems that can be considered WD J0914-like would have to start out at inclinations around $50\degree$ or $130\degree$, with high $\epsilon$ values above $0.02$, in order to reach close orbits or undergo planetary tidal disruption within the $\sim13$~Myrs observed for WD J0914.

\begin{figure*}
\hspace{0.0\linewidth}
\includegraphics[width=\linewidth]{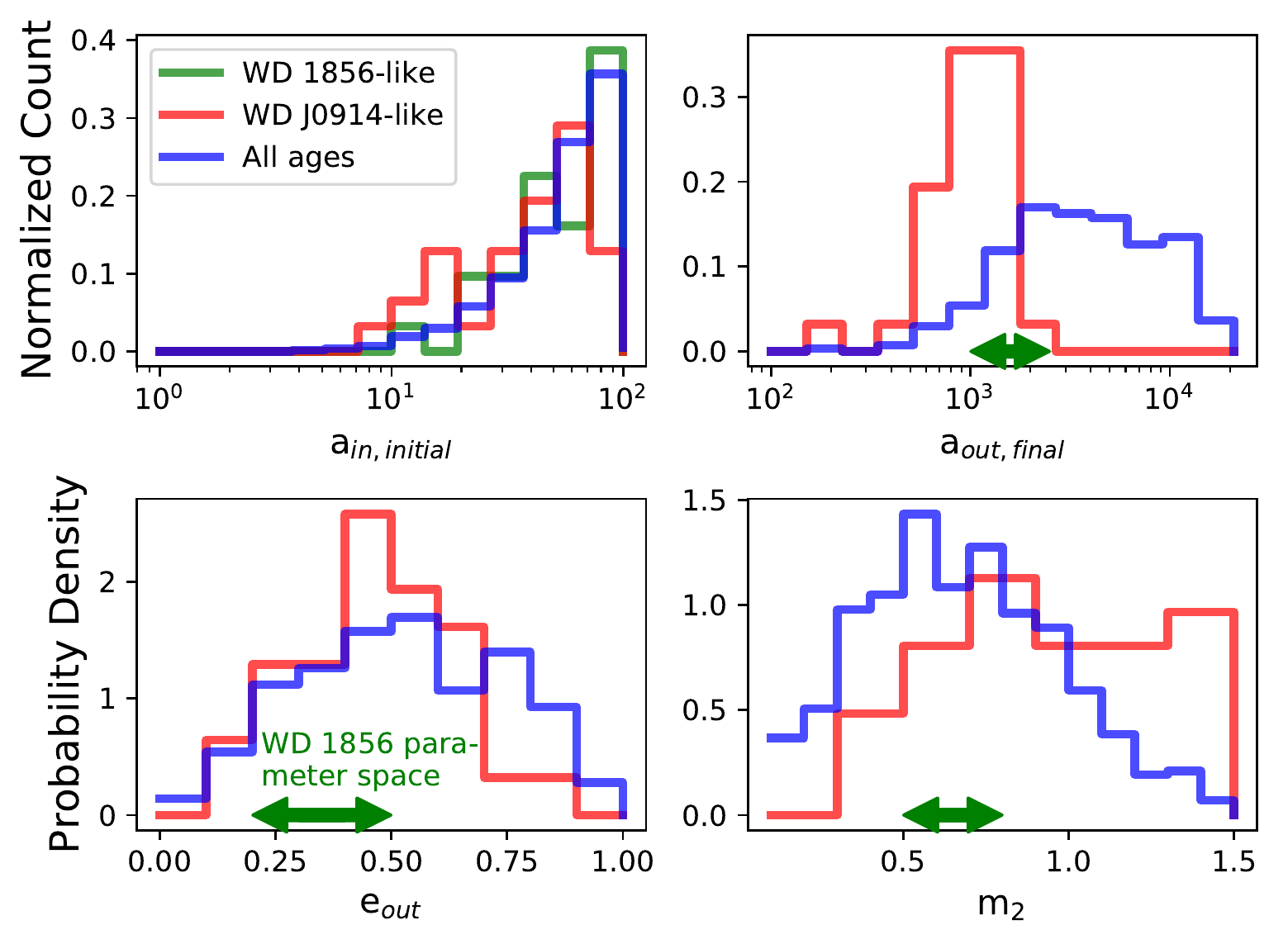}
\caption{{{\bf Orbital parameter distributions for producing WD polluters and short period planets consistent with the WD 1856 and WD J0914 systems.} The four panels in this figure show the distributions of the orbital parameters that can lead to the formation of short period planets around WDs or to planets polluting WDs. The upper left panel shows the distributions for the initial planet SMA values, the upper right panel shows the distributions for the final stellar companion SMA values, the lower left panel shows the distributions for the stellar companion eccentricity values, and the lower right panel shows the distributions for the stellar companion masses. In each panel, the red histograms show the distributions for systems that are consistent with WD J0914, the green histogram and arrows show the distribution or part of the parameter space that is consistent with WD 1856, and the blue histograms show the distributions for all WD cooling ages (for comparison).} }

\label{fig:hists}
\end{figure*}

Beyond the inclination-$\epsilon$ space, the results also give us restrictions for the initial planetary SMA, final stellar companion SMA, as well as the stellar companion eccentricity and mass that facilitate the migration or disruption of planets. As can be seen in Fig. \ref{fig:hists}, upper left panel, the initial planet SMA values should be larger than about $10$~AU, with the maximum likelihood at our largest tested SMA of $100$~AU, considering all cooling ages (blue histogram). {While initial SMAs smaller than $10$~AU are possible, they are much less likely based on our results.} While it could be expected form this result that the likelihood would increase even further for some larger SMA values, we do not include such systems as observations currently do not support such large planet SMA values in abundance \citep[e.g.,][]{Fernandes+2019}. Note that tidal migration from wider initial SMAs may lead to more significant tidal disruption of these planets, reducing the strong preference for wider initial SMAs in our results somewhat. When only considering those systems where the planet migration occurred rapidly enough to reproduce WD J0914 (on the order of $\sim20$~Myrs\footnote{While the estimated cooling age of WD J0914 is only $13$~Myrs, we include results up to $\sim20$~Myrs for better statistics as the formation mechanism is consistent.}, red histogram), the distribution for initial planet SMAs still favors larger values, though less strongly, putting the average closer to the low tens of AU. The planet SMA values for systems that follow the WD 1856 restrictions listed in Sec. \ref{sec:setup} (green histogram)  broadly follow the general planet SMA distribution for all cooling ages.

In Figure \ref{fig:hists}, upper right panel, we show the distribution of final stellar companion SMA values for systems that reach close configurations within $20$~Myrs of the primary becoming a WD (red histogram) and for all WD cooling ages (blue histogram). Here, we see a clear distinction between these two groups of cooling ages, such that the systems evolving faster require stellar companions between SMA values of $\sim500$ and $2000$~AU, while for all cooling ages, the majority of stellar companions will be between $\sim1000$ and $10000$~AU (disregarding SMA values wider than that as they would be unstable against galactic tides). We highlighted the companion SMA's estimated value for WD 1856 with a green double arrow, demonstrating that it is consistent with our results. 

In Figure \ref{fig:hists}, lower left panel, we show the results for the stellar companion orbital eccentricity distributions, using the same meanings of colors for the histograms as in the previous figures. For both the $20$~Myrs and all cooling ages cases, very small and very large companion eccentricities are disfavored, with both distributions peaking around the value of $0.5$. Again, the results appear fully consistent with the estimates for WD 1856's companion from \citet{Vanderburg+2020}.

Finally, in Figure \ref{fig:hists}, lower right panel, we show the distributions of stellar companion masses for the two cases. For all cooling ages, a companion mass larger than $0.3~M_\odot$ and smaller than $1.0$~M$_\odot$ is preferred, with a peak around $0.6$~M$_\odot$ (consistent with WD 1856), while for the $20$~Myrs case larger stellar masses are not disfavored. However, in general, the range for possible companion masses is quite large.

\section{Discussion and Conclusion}\label{sec:disc}

The recent discoveries of WD J0914 and WD 1856 have reaffirmed the notion that planets exist around stars of all ages and evolutionary stages. Furthermore, as the planets in both systems could not have existed at their current orbits during their hosts' earlier life stages, the observed configurations indicate a complex dynamical evolution for these systems. WD J0914's short cooling age additionally indicates that any dynamical mechanism in question must be able to act very fast after the host had become a WD. Here, we have demonstrated that the EKL mechanism, combined with the effects of stellar mass loss, can naturally explain systems like WD J0914, as well as systems like WD 1856.

However, a few caveats exist that have to be considered when applying our results on either WD J0914 or WD 1856. For WD J0914-like systems, a key element is that tidal migration has to occur fast and, in order to reproduce the estimated current planet SMA value of $\sim0.07$~AU, has to be more effective than ``regular'' tidal dissipation mechanisms would allow. A possible solution for these issues are models of chaotic tides, which have been invoked to explain the rapid migration \citep{VerasFuller2019a,VerasFuller2019b}. Furthermore, WD J0914 does not have a known stellar binary companion; indeed, our work predicts the possible parameters of a potential companion that might have been missed by previous observations. 

For WD 1856-like systems, the fact that its distant companion is a stellar binary itself may introduce additional dynamical effects \citep{Hamers+2015, HamersPZ2016, HamersLai2017, OConnor+2020}. \citet{OConnor+2020} estimated that these effects can significantly enhance the formation of high eccentricity orbits, for specific orbital configurations. In particular, they estimated that WD 1856 b had an initial orbit between $10$ and $20$~AU to make effective use of these four-body effects.

\citet{Munoz+2020} also investigated the dynamics of WD 1856, not considering the potential additional dynamical effects from the companion binary, but only the KL effect itself. They deduce, based on analytical considerations of planet survival to the WD phase and the parameter space that allows large eccentricity excitations, but also excluding planets that cross the Roche limit, that the initial planet SMA would have been $\sim2-2.5$~AU\footnote{We note that the \citet{Munoz+2020} parameter space appears to be more restricted compared to our findings. The source of this difference can be traced to their condition that the minimum periapsis distance due to EKL oscillations must be larger than the size of the star during the red giant phase. In reality, the eccentricity oscillations in a given system can have a longer period than the duration of the red giant phase, such that the entire red giant phase has a shorter duration than the EKL mechanism timescale. In other words, the timescale between the two high-eccentricity peaks that would have brought the planet into the stellar envelope may be longer than the entire giant phase. Our numerical analysis here allows us to explore such cases.}. 

Given these caveats, our results still allow us to make predictions for the orbital parameters of both WD J0914 and WD 1856, and to compare them to the predictions of other works.

\subsection{Implications for WD J0914}\label{sec:imp1}

For a planet to reach high enough eccentricity values quickly enough after WD formation to be consistent with the observations of WD J0914, several orbital parameters have to have been within fairly tight bounds. The initial inclination of the planet's orbit must have been close to either $50\degree$ or $130\degree$ (see Fig. \ref{fig:Paramspace}, and the current SMA value of the predicted stellar companion must be in the range of $\sim500$ to $3000$~AU. On the other hand, the stellar companion's eccentricity is not well constrained; other than that, it should be close to neither zero nor $1$. Likewise, the planet's initial SMA is relatively loosely constrained, with plausible values reaching from $10$ to $100$~AU. The stellar companion mass is also not well constrained, allowing a broad range of stellar types, as long as the mass is larger than $\sim0.3$~M$_\odot$.

{We searched for bound companions to WD J0914, restricting ourselves to companions with angular separations of $<15"$. WD J0914 has a Gaia eDR3 parallax of $\pi=2.34\pm 0.29~{\rm mas}$ \citep{Gaia:2021}, corresponding to a distance of $d=427\pm 52~{\rm pc}$. We note that \citet{Bailer-Jones:2021} find a distance of $d=475~{\rm pc}$; our qualitative conclusions would be unchanged if we were to adopt this distance.  Thus, at the distance of WD J091, angular separations of $<15"$ correspond to projected separations of $\la 6400$~AU, fully encompassing the likely range of final SMA for a companion as shown in Figure~\ref{fig:Paramspace}.  Querying {\it VizieR}\footnote{\url{https://vizier.u-strasbg.fr/viz-bin/VizieR}}, we find two sources in this region: WD J0914 itself, and a companion $\sim 13"$ away that is $\sim3$ magnitudes brighter than WD J0914 in $G$.  However, this companion is not bound to WD J0914, as its parallax and proper motion differ from WD J0914 by $\sim 4\sigma$ (parallax), $\sim11\sigma$ (proper motion R.A.), and $\sim 12\sigma$ (proper motion Dec.). These differences cannot be explained by a line-of-sight separation or orbital motion of a bound companion.  Rather, the second source is likely a foreground K3V or K4V star.}

{We can place an upper limit on the mass of any putative stellar or brown dwarf companion to WD J0914 using the UKIRT Hemisphere Survey (UHS; \citep{Dye:2018}), which reports a median $5\sigma$ point source sensitivity of $J=19.6$ (Vega).  Figure 2 of \citet{Dye:2018} indicates that the sensitivity is spatially variable over the survey area at the $\sim 0.5$ magnitude level.  We queried the UHS catalog\footnote{\url{http://wsa.roe.ac.uk:8080/wsa/region_form.jsp}} for sources within 5 arcminutes of WD J0914. We find that all sources have $J<19.6$ (using jAperMag3 magnitudes) or have $J$ magnitudes that are consistent with this limit at $1\sigma$.  We therefore adopt J=19.6 as the sensitivity limit. At the distance of WD J0194 ($d=381-487~{\rm pc}$), this corresponds to an absolute magnitude of $M_J=11.2-11.7$, assuming that the extinction in $J$ is negligible.  From Table 4 of \citet{Hoard:2007}, this corresponds to a star at the bottom of the main sequence with a spectral type of M8-L0.  Using the models of \citep{Baraffe:1998}, these absolute magnitudes correspond to stars of mass $\sim 0.075-0.086~M_\odot$ for ages of 1-10 Gyr. We note that this assumes that the companion is a single star. However, the companion itself could be a binary (similar to the WD 1856 system), which would allow for a larger total companion mass, as the individual stars would be less massive and thus less luminous.}

{Thus we find that companions with separations and masses within the most likely range needed to emplace the polluting companion to WD J0914 ($a_{\rm out,final}=500-3000$~AU and $m_2>0.3~M_\odot$) are ruled out.  For completeness, we note that \citet{Gaensike+2019} placed a more stringent constraint on the spectral type of any putative bound companion to WD J0914 of L5.  This is because they assumed the companion was coincident with WD J0914, which they estimate has a magnitude of $J=19.65$ based on their synthetic spectrum.  WD J0914 is not detected in the UHS, but assume their estimate of its magnitude of $J=19.65$, must be just barely below the detection limit of $J_19.6$. Thus one can place a 5$\sigma$ upper limit on the magnitude of any coincident companion of $J\simeq 23$, or an absolute magnitude of $M_J\simeq14.6$ assuming no extinction and a conservative $1\sigma$ upper limit to the eDR3 distance of $d=487$~pc.  This corresponds to brown dwarfs with spectral type L8 and mass $\sim 0.07~M_\odot$ \citep{Baraffe:1998,Hoard:2007}.  However, our simulations indicate that any companion responsible for the emplacement of the polluter of WD J0914 is likely to be separated by $>500$~AU or $>1"$, and thus our upper limit is likely more relevant. }

Within our models we considered both systems that lead to high-eccentricity migration, as well as systems that lead to Roche-limit crossing of the planet as potential pathways to form WD J0914. This is in part motivated by the observed accretion disk around the WD that may be a result of tidal stripping of the planet, making the currently suspected planet in the system a remnant of a larger progenitor. An immediate consequence of such a formation scenario is that the WD did not just absorb a significant amount of planetary material, but also angular momentum. This extra angular momentum would have spun up the star to fast rotation speeds, potentially as fast as hours or even minutes \citep{Stephan+2020}. A recently discovered rapidly rotating WD, SDSS J121929.45+471522.8, which is suspected to have a planetary companion as well based on the characteristics of its magnetic field, may indeed be an example of such a spin-up due to the consumption of planetary material and angular momentum \citep{Gaensicke+2020}.

\subsection{Implications for WD 1856}\label{sec:imp2}

Since WD 1856 has quite a large cooling age, a broader set of initial conditions can reproduce such a system. However, since a companion binary star is indeed known to exist, with some estimates for its orbital parameters, we have to restrict ourselves to only systems consistent with those constraints. As such, the initial inclination of the system could be nearly anywhere between $40\degree$ and $140\degree$, and the initial planet SMA value was most likely in the range of $\sim10$ to $100$~AU, with a preference to larger values. This contrasts to the predictions of both \citet{OConnor+2020} and \citet{Munoz+2020}, which are more restricted. Indeed, the different predictions for the initial planet SMA might translate into different predictions for the planet's formation, composition, and history. 

In general, a more distant formation site might lead to the inclusion of more volatile chemical species, such as ammonia or methane, or to the formation of a more massive planet. The presence of such molecules might be detectable through transit spectroscopy with the James Webb Space Telescope (JWST), which, given the transit depth and estimated scale height of the planet, could have an easily detectable signal \citep{Vanderburg+2020}. However, the heat deposited into the planet during tidal migration, as it is the primary site of tidal dissipation in this system, may have destroyed any complex chemistry. The mass estimate by \citet{Vanderburg+2020} is fairly uncertain given the current data, however, the high end of the mass estimate ($\sim12$~M$_{Jup}$) could support the idea of a distant planet formation site, potentially more consistent with brown dwarf-like formation than planetary formation models. However, like for WD J0914, we estimate that the currently observed planet could also be a remnant planet of a larger, partially tidally disrupted, progenitor. In this case we again would expect the WD to rotate very rapidly compared to regular WDs.

\subsection{Conclusion}\label{sec:sum}

In summary, the mass-loss-enhanced EKL mechanism naturally explains the formation of short period planets around WDs, as well as WD pollution by a variety of planet types. This mechanism is able to produce both WD J0914 and WD 1856-like systems, {however, our model predicts that WD J0914 would have a low-mass stellar companion on an orbit with an SMA value between $\sim500$ and $3000$~AU, which seems to be excluded by observations. The companion would thus either need to be a binary of very low-luminosity red or brown dwarfs, or an older, cooled WD. In the latter scenario, the companion star would have been the more massive binary member initially, evolving past the main sequence faster than the planet's host star. While such a scenario is possible, it introduces additional (although tractable) complexity in the numerical model due to the consecutive mass loss phases of the member of the original binary.  A thorough exploration of such scenarios is interesting, but goes beyond the scope of this current work.} We predict that WD 1856 b's initial SMA value was most likely between $10$ and $100$~AU, in contrast to other models. The different predictions might be disentangled with future observations of the planetary chemistry through transit spectroscopy with JWST, better estimates of the planetary masses, as well as measurements of the WDs' rotation speeds.

\section*{Acknowledgements}
We thank Chris Kochanek and Kevin Luhman for helpful discussions, and the anonymous referee for comments and suggestions that improved the paper. A.P.S. acknowledges partial support from a President's Postdoctoral Scholarship from the Ohio State University, and the Ohio Eminent Scholar Endowment. 
A.P.S. and B.S.G. acknowledge partial support by the Thomas Jefferson Chair Endowment for Discovery And Space Exploration. A.P.S. and S.N. acknowledge partial support from the NSF through grant No. AST- 1739160. S.N. thanks Howard and Astrid Preston for their generous support.  This research has made use of the SIMBAD and VizieR databases, operated at the CDS, Strasbourg, France, and NASA’s Astrophysics Data System Abstract Service. This work has made use of data from the European Space Agency (ESA) mission Gaia (\url{http://www.cosmos.esa.int/gaia}), processed by the Gaia Data Processing and Analysis Consortium (DPAC,
\url{http://www.cosmos.esa.int/web/gaia/dpac/} consortium). Funding for the DPAC has been provided by national institutions, in particular the institutions participating in the Gaia Multilateral Agreement.



\bibliographystyle{aasjournal}
\bibliography{Kozai2} 

\end{document}